\documentclass[aps,prb,twocolumn,groupedaddress,amsmath,amssymb,showpacs]{revtex4-1}

\usepackage{graphicx,graphics,color,epsfig}
\usepackage{amsmath}
\usepackage{amssymb}
\usepackage{amsfonts}
\usepackage{amsfonts}
\usepackage{epstopdf}
\usepackage{bm}
\usepackage{times,xspace}
\usepackage{color}
\usepackage[utf8]{inputenc}

\def\beq{\begin{eqnarray}}\def\eeq{\end{eqnarray}}
\def\be{\begin{equation}}\def\ee{\end{equation}}

\def\l{\lambda}

\begin{document}

\title{Photo-induced SU(3) topological material of spinless fermions}

\author{Sayonee Ray\footnote{sayoneeiisc@gmail.com}, Ananya Ghatak\footnote{gananya04@gmail.com} and Tanmoy Das\footnote{tnmydas@gmail.com}\\
{Department of Physics, Indian Institute of Science, Bangalore-560012, India.\\
}}
%\address{Email: tnmydas@gmail.com.}

\date{\today}

\begin{abstract}
Generation of topological phases of matter with SU(3) symmetry in a condensed matter setup is challenging due to the lack of an intrinsic three-fold chirality of quasiparticles. We uncover two salient ingredients required to express a three-component lattice Hamiltonian in a SU(3) format with non-trivial topological invariant. We find that all three SU(3) components must be entangled via a gauge field, with opposite chirality between any two components, and there must be band inversions between all {\it three} components in a given eigenstate. For spinless particles, we show that such chiral states can be obtained in a tripartite lattice with three inequivalent lattice sites in which the Bloch phase associated with the nearest neighbor hopping acts as $k$-space gauge field. The second and a more crucial criterion is that there must also be an odd-parity Zeeman-like term, i.e. $\sin(k)\sigma_z$ term where $\sigma_z$ is the third Pauli matrix defined in any two components of the SU(3) basis. Solving the electron-photon interaction term in a periodic potential with a modified tight-binding model, we show that such a term can be engineered with site-selective photon polarization. Such site selective polarization can be obtained in multiple ways, such as using Sisyphus cooling technique, polarizer plates, etc. With the $k$-resolved Berry curvature formalism, we delineate the relationship between the SU(3) chirality, band inversion, and $k$-space monopoles, governing finite Chern number without breaking the time-reversal symmetry. The topological phase is affirmed by edge state calculation, obeying the bulk-boundary correspondence.
\end{abstract}
\maketitle

\section{Introduction}
The discovery of quantum Hall effect in 1980~\cite{klitzing} has ushered in a new era of quantum states of non-interacting electrons, distinguished by non-trivial topological invariant. Subsequently, Haldane's proposal\cite{haldane} of quantum Hall effect with synthetic gauge field in a honeycomb lattice gave an important clue on how a specific lattice structure can give rise to non-trivial band topologies. There are various mechanism by which the momentum-space mapping of the Hamiltonian with specific crystal symmetry, orbital symmetry, and/or spin-orbit coupling (SOC) results in an irreducible SU(2) representation of the Dirac or Weyl Hamiltonian, garnering a large class of Dirac and Weyl materials.~\cite{ssh_rmp,TIreviewTD,TIreviewCK,TIreviewSCZ,castroneto,Balatsky,Ando,Vishwanath} In bipartite lattices, such as in the so-called Su-Schrieffer-Heeger model,\cite{ssh_rmp} or in honeycomb lattice,\cite{castroneto} the hopping between inequivalent lattice sites carries a net Bloch phase $e^{i{\bf k}\cdot{\bf r}}$. This acts as a momentum-space gauge field, in analogy with the Wilson loop formula for a magnetic field which, in special cases, is associated with an integer phase winding number $-$ a non-trivial topological invariant. Again, electron hopping between even and odd parity orbitals (such as $s$- and $p$-orbitals) often results in a net hopping $\propto e^{i{\bf k}\cdot{\bf r}}-e^{-i{\bf k}\cdot{\bf r}}=2i\sin{{\bf k}\cdot{\bf r}}$, giving a linear orbital-momentum locking.\cite{WeylTD} As orbital texture inversion occurs at discrete $k$-points, these points act as monopoles in the momentum-space, giving rise to Dirac or Weyl cones. Furthermore, SOC provides a common origin to a variety of quantum Hall and topological classes of materials.\cite{TIreviewTD,TIreviewCK,TIreviewSCZ} In all these examples, the spinor of the SU(2) representation (or its generation to a SU(2$N$) version, where $N$ is integer) comprises of two chiral species originating from the momentum locking with two sublattices, or two orbitals, or spin-1/2 particles. As a chiral object forms an orbit, surrounding an external magnetic field, or momentum space Berry curvature, it produces a non-zero flux, which is quantified by the winding number or Chern number (topological invariants). 
 
Encouraged by the tremendous success of material realization and engineering of the SU(2) based topological materials,\cite{TIreviewTD,Ando,EngineeringTI} we explore the possibility of designing a lattice model with SU(3) symmetry. SU(3) flavor symmetry was originally proposed in the quantum chromodynamics (QCD) field for systems with three basis vectors (namely, colors) which can be expanded in the basis of eight $3\times 3$ Gell-Mann matrices $\hat{{\bm \l}}$.~\cite{mukunda,georgi,gellmann} The non-Abelian theory of the SU(3) groups predicts several elementary excitations such as quark, and gluons. Prediction of emergent SU(3) symmetry in condensed matter systems is rather limited,\cite{galitski,jpsj} and no physical system or optical lattice is realized to date with such properties. In a recent work, it is shown that a SOC generated in a square optical lattice in the presence of a spatially homogeneous SU(3) gauge field can give rise to non-trivial topological characteristics ~\cite{galitski}. This can be experimentally realized with ultra-cold atoms having internal spin degrees of freedom or any such three component Hamiltonian. An explicit topology-engineering scheme in a three band model has been discussed in another recent work~\cite{jpsj},  where the generation of arbitrary Chern number is based on the equivalence of the topological number of a given band on the monopole charge, and can be extended to higher Chern number derivatives. The feasibility of the materials realization of the corresponding models has not been discussed in these papers.

The basic principles for SU(3) symmetric TI follows that of the SU(2) counterpart. A momentum representation of the SU(3) Hamiltonian incipiently demands the entanglement of a three components quantum state via some sort of intrinsic gauge field. We start with a generalized form of the SU(3) Hamiltonian $\hat{H}({\bf k})={\bf b}({\bf k})\cdot \hat{{\bm \l}}$, where ${\bf b}({\bf k})$ is the eight components vector made of electron hoppings in a lattice. The momentum-space magnetic field (Berry curvature) can thus be expressed in (2+1) dimension,\cite{galitski}
\begin{equation}\label{Berry}
\Omega_{mn}({\bf k}) = \frac{1}{2|{\bf b}|^3}\epsilon^{\mu\nu\rho}b_{\mu}\partial_{m}b_{\nu}\partial_{n}b_{\rho},
%b_{\nu}b_{\rho} %\frac{1}{2|{\bf b}({\bf k})|^3} {\bf b}({\bf k})\cdot \partial_{k_x} {\bf b}({\bf k}) \times \partial_{k_x} {\bf b}({\bf k}).
%
\end{equation}
where indices $m$, $n$ = $k_x,~k_y$, and $\mu$, $\nu$, $\rho$ give the components of the ${\bf b}$-vector. $\epsilon^{\mu\nu\rho}$ is the usual Levi-Civita tensor.  Eq.~\eqref{Berry} implies that chiral orbits or vortices are formed in all three bands, with their centers located at $|{\bf b}({\bf k}^*)|=0$, where the Berry curvature $\Omega$ obtains singularity. The flux of $\Omega({\bf k}^*)$ through the first Brillouin zone for each band is quantized, and is quantified by the associated Chern number. Therefore, the Chern number essentially dictates the number of orbit centers, while its sign corresponds to the direction of associated phase winding or chirality. Without any external magnetic field, the total Chern number for all bands must vanish. This implies that the chirality of one of the SU(3) component must be opposite to the other two components, and also the associated Chern number (or $k-$space monopoles), say, $C_1$ must be equal to the total Chern number from the other two chiral states, i.e. $C_1=C_2+C_3$. In the case of $C_1=0$, the other two bands give $C_2=-C_3$ which mimics the quantum spin-Hall system for spinful systems.\cite{TIreviewCK,TIreviewSCZ}

The above-mentioned features can also be understood simply by the corresponding band topology. The center of orbits at ${\bf k}^*$-points are those discrete points where band degeneracy occurs. Therefore, the number of vortices dictate the number of band inversions in two dimensions (2D). In SU(2) topological systems, the odd number of band inversions (at the time-reversal momenta, if this symmetry is present) between two basis components gives a finite Chern number or $Z_2$ invariant. For SU(3) systems, the band inversion must happen between all three bands, or at least, the band with the highest Chern number ($C_1$) must undergo inversion with {\it both} the other two bands. The other two bands do not necessarily undergo a band inversion between them unless their Chern numbers are also different. This is an important distinction of the SU(3) framework. Another unique requirement of the SU(3) topological state is that here not only a gauge field is required to be present in the off-diagonal term of the Hamiltonian (as in SU(2) case), but {\it an odd parity Zeeman-like term, i.e. $\sin{k}\sigma_z$ term must also be present between any two basis}. Such an odd parity onsite Zeeman-like term does not arise naturally from Bloch phase or from conventional SOC. 

\begin{figure}
%\hspace{-2.2cm}
\centering
\includegraphics[width=0.8\columnwidth]{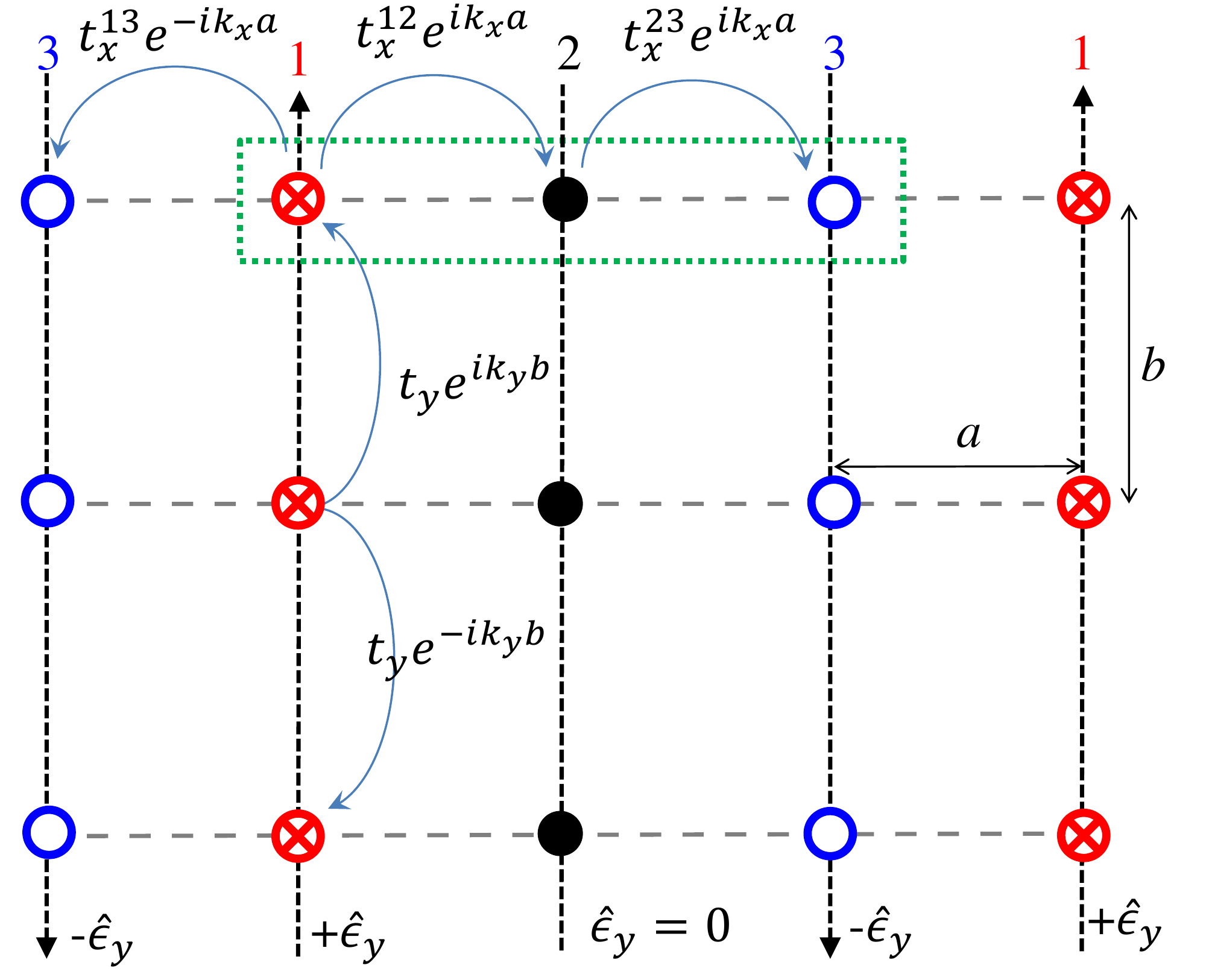}
\caption{(Color online) Schematic drawing of the proposed setup. The different coloured spheres denote three different basis of the Hamiltonian, which can be three different atomic species or orbitals, $\psi_1$, $\psi_2$ and $\psi_3$. ${\bf E}$ denotes the direction of the vector potential of the electromagnetic (EM) wave, which causes the different Peierls phase coupling for the different orbitals. The hopping amplitudes are multiples of $t_x$ and $t_y$ , and are different for the different components (orbitals). Also, $b=2 a$ in the present model.
}
\label{model}
\end{figure}

A focal point of our work is to construct a lattice model which inherits the required suitable gauge fields, and odd-parity Zeeman term, and thus intrinsically performs as topological material distinguished by finite Chern number. We present a model Hamiltonian of a spinless SU(3) topological system in a tripartite lattice with three inequivalent lattice sites. Each sublattice is sitting in distinct 1D chains, and they are coupled by nearest-neighbor quantum tunneling, as shown in Fig.~1 . This would naturally give an uncompensated Bloch phase for all three inter-site hoppings as: $e^{i{\bf k}.{\bf r}_{12}}$, $e^{i{\bf k}.{\bf r}_{23}}$, while the third one is naturally reversed to $e^{-i{\bf k}.{\bf r}_{13}}$, where ${\bf r}_{ij}$ are distances between $i$ and $j$ sublattice. This naturally enforces opposite chirality between any two band in a given parameter space. Finally, for the generation of the odd parity `onsite' matrix-element, we propose to utilize the light-matter interaction with site-selective photon polarization. We show with a modified tight-binding model in a tripartite lattice that the dipole interaction term $\propto {\bf k}\cdot{\bf A}$, where ${\bf A}$ is the vector potential of the photon, naturally leads to a $\pm\sin{(\beta {\bf k}\cdot\hat{\epsilon})}$ term (where $\hat{\epsilon}$ is the unit vector along the polarization direction, and $\beta$ is a tunable parameter). Then by tuning the direction of the polarization parallel or orthogonal to ${\bf k}$ (or with other methods as discussed later), one can selectively generate, reverse, and destroy this term in different sublattices. This way, we can generate odd-parity Zeeman term by using opposite polarization in lattice sites 1 and 2, while it is destroyed in the 3rd site, as shown in Fig.~1. This generates our SU(3) topological system, which we characterize by the detailed analysis of the Berry curvature, Chern number, and edge states. We also discuss two realizable setup to obtain site-selective polarization with existing tools. 

The rest of the paper is arranged as follows. In Sec.~\ref{Sec:II}, we describe the general characteristics that the SU(3) Hamiltonian should possess to obtain non-zero Chern number. In Sec.~\ref{model_build} we discuss the setup. Derivation of the odd-parity Zeeman term with site-selective EM fields is given in Sec.~\ref{Sec:EPcoupling}. The tight-binding model for a triparticle lattice is discussed in Sec.~\ref{Sec:lattice}. Design principles for such lattice with Sisyphus cooling technique or with polarizer plate are given in Sec.~\ref{Sec:design}. In Sec.~\ref{cal_chern} we elaborate on the geometrical method for calculating the Berry curvature and Chern number. We detail the calculation of edge states using the strip geometry approach in Sec.~\ref{cal_edge}. In Sec.~\ref{spin} we discuss the robustness of the spinless SU(3) topological phases to spinful perturbations. We also discussed a possible model of engineering SU(3) by suitably combining SU(2) and U(1) species. We end the paper with discussions and conclusions in Sec.~\ref{conclusion}. 

\section{Model}\label{Sec:II}

\subsection{General characteristics of SU(3) TI Hamiltonians}\label{gen_char}
We design a SU(3) Hamiltonian with on eye on finite Chern number in a bottom-up approach. A generic Hamiltonian, obeying the SU(3) decomposition, can be written as,
\be \label{su3_gellmann}
\hat{H}({\bf k})=a({\bf k})\hat{{\mathbb{I}}}_3+{\bf b}({\bf k})\cdot {\bf \hat{\l}}\,,
\ee
where $\hat{{\mathbb{I}}}_3$ is a 3$\times$3 identity matrix, $\hat{\l}_i$ are the SU(3) generators (Gell-Mann matrices), and $a({\bf k})$, $b_i({\bf k})$ are the corresponding coefficients. The explicit matrix form of the Hamiltonian (the $k$- dependencies in $a$ and $b_i$ are implied) is,
\be\label{gensu3}
\hat{H}({\bf k})=\begin{pmatrix}
a+b_3+\frac{b_8}{\sqrt{3}}& b_1-ib_2& b_4-i b_5\\
b_1+i b_2& a-b_3+\frac{b_8}{\sqrt{3}}& b_6-i b_7\\
b_4+i b_5& b_6+i b_7& a-\frac{2b_8}{\sqrt{3}}\\
\end{pmatrix}.
\ee
We work with a three component spinor $\Psi_k^{\dag}=\left( {\psi}_1^{\dag}(k), {\psi}^{\dag}_2(k),{\psi}^{\dag}_3(k)\right)^{T}$, where $\psi_i(k)$ are the basis  representing different orbitals, or sublattices, and so on (but we do not consider spin here). Each $b_i$ term requires special treatment such that opposite chirality, and odd-parity Zeeman term can be simultaneously achieved in such a way that Berry curvature singularities at discrete $k$-points can be attained.  

\subsubsection{Diagonal terms} \label{diag_terms}
We start with the diagonal terms of Eq.~\eqref{gensu3}. We denote the three onsite, intra-basis, dispersions as $\xi_{i}({\bf k})$ where $i=1,~2,~3$. In general tight-binding Hamiltonians, diagonal terms comprise of cosine functions of momentum, and chemical potential. The sine term of the Bloch phase is associated with imaginary `i' which cannot appear in the diagonal term for it to be a Hermitian one. It drops out even in centrosymmetric lattices as the same hopping on both positive and negative directions are added. With the analysis of Berry curvature and SU(3) symmetry, we recognize that an essential requirement for non-zero Chern number in this case is that the diagonal terms $\xi_1$ and $\xi_3$ must contain $\pm \sin(k_y)$ terms, which is equivalent to having odd-parity Zeeman term. Without specifying the origin at this point, we start with a combination of three diagonal terms in a 1D lattice:
\beq\label{diagonal}
\xi_i({\bf k}) &=& t_i\cos{(\alpha k_y)}+ m_i\sin{(\beta k_y)}-\mu, 
\eeq
%\xi_1({\bf k}) &=&  2t_y\cos{(\alpha k_y)}-m_1\sin{(\beta k_y)}-\mu, \nonumber\\
%
%\xi_2({\bf k}) &=&  -\frac{3}{2} t_y \cos{(\alpha k_y)}-\mu, \nonumber\\
%
%\xi_3({\bf k}) &=&  t_y\cos{(\alpha k_y)} + m_2 \sin{(\beta k_y)}-\mu,
%\eeq
%
where $t_i$, $m_i$ are the expansion parameters, and $\mu$ is the chemical potential. $\alpha$ and $\beta$ are arbitrary parameters depending on the crystal structure and lattice constants. Finite Chern number arises for a set of parameters as $t_1=-2t_y$, $t_2=\frac{3}{2}t_y$, $t_3=-t_y$, and $m_1=-m_3$, and $m_2=0$. The cosine terms arise from the nearest neighbor hopping along the $y$-direction. In Sec.~\ref{Sec:EPcoupling} below, we discuss how to obtain $m_i\sin{k_y}$ term with the help of light-matter interaction in which we find that $m_i$ depends on both $t_i$ as well as the vector potential $A$. Thus its sign can be simultaneously reversed by using antiparallel photon polarization. We notice that all three diagonal terms are taken to depend only on $k_y$ which is consistent with the setup drawn in Fig.~1.  By comparing Eqs.~\eqref{diagonal} and \eqref{gensu3}, we obtain
\beq\label{ab3b8}
a({\bf k}) &=& \frac{\xi_1+\xi_2+\xi_3}{3} = \frac{1}{6} \big(-3 \cos{\alpha k_y} + 2 (m_1 + m_3) \sin{\beta k_y} \big), \nonumber\\
b_3({\bf k}) &=& \frac{\xi_1-\xi_2}{2} = \frac{1}{4}(-7 \cos \alpha k_y  +2 m_1 \sin\beta k_y), \nonumber\\
b_8({\bf k}) &=& \frac{2\xi_3-\xi_1-\xi_2}{2\sqrt{3}} \nonumber\\
&=& \frac{1}{4\sqrt{3}}(3 \cos\alpha k_y + 2(m_1 - 2 m_3) \sin\beta k_y), %\nonumber\\
\eeq
Looking at the Hamiltonian in Eq.~\eqref{gensu3}, we notice that $a({\bf k})$ gives  a overall shift to all the bands and thus does not play any specific role on the topology. $b_3({\bf k})$ gives an anisotropic Zeeman splitting between 1st and 2nd basis in the Hamiltonian, while $b_8({\bf k})$ gives a similar splitting of the 3rd basis from the other two ones. It is easy to see that the band inversion along the $k_y$ direction is driven by $b_3$ and $b_8$ terms. And also, since the eigenvalues are proportional to $b_3$ and $b_8$, we see that the bands become anisotropic between $\pm k_y$. On the other hand, along the $k_x$ direction they are symmetric, since the eigenvalues depend on the absolute value of the other $b_m$ terms. This asymmetry also reflects in the Berry curvature maps shown in Fig.~\ref{3Dbren}.

\subsubsection{Off-diagonal terms} \label{off_diag}
Next we consider the three off-diagonal terms which follow a general form $b_{\nu}({\bf k}) \pm i b_{\sigma}({\bf k})$, where $\nu=1, 4, 6$, $\sigma=2, 5, 7$. In condensed matter systems, such a complex term usually has two origins: (1) Rashba- or Dresselhaus-type spin-orbit coupling (SOC), (2) Bloch phase from nearest neighbor electron's hopping. (1) Rashba and Dressenhaus SOC yields $b_{\nu}({\bf k})=\alpha_R \sin{k_x}$, and $b_{\sigma}({\bf k})=\alpha_R\sin{k_y}$ (where $\alpha_R$ is the SOC strength). SOC is however difficult to achieve for all three SU(3) spins in both condensed matter and optical lattice setups. More importantly, we find that the computation of Chern number with SOC in the off-diagonal terms often gives zero Chern number. Therefore, we focus on the possibility (2). Assigning $b_{\nu}({\bf k})=t_x\cos{k_x}$, and $b_{\sigma}({\bf k})=t_x\sin{k_x}$ (where $t_x$ is a parameter which can be different for different $\nu$ and $\sigma$), we see that this term simplifies to $\sim t_x\exp(ik_x)$. This is just a Bloch phase associated with the nearest-neighbor hopping between different sublattices (since it appears in the off-diagonal term in the Hamiltonian). We also find that for finite Chern number, {\it the Bloch phase must be reversed in at least one of the off-diagonal terms, compared to the other two.} 

\subsubsection{Full Hamiltonian}
Based on the constraints for both diagonal and off-diagonal terms, we now seek a minimal model for the realization of SU(3) Chern insulator in the spinless basis:
\begin{eqnarray}
H(k)=\left(
 \begin{array}{ c c c c }
\xi_1(k_y) & -t_x e^{ik_x} &  -t_x e^{-ik_x}    \\
-t_x e^{-ik_x}  & \xi_2(k_y)  & - \frac{1}{2}t_x\ e^{i k_x}    \\
-t_x e^{ik_x} & -\frac{1}{2}t_x\ e^{-i k_x} &  \xi_3( k_y)\\
\end{array} \right),
\label{Ham1}
\end{eqnarray}
where $t_x$ is the nearest neighbor tight-binding hopping parameter between different basis. Without loosing generality, we set $t_x=t_y$=1. This gives all eight components of the ${\bf b}$ vector to be:
\begin{align}
\begin{split} 
{\bf b}(k) &= \big[-\cos k_x, -\sin k_x, 
 \frac{1}{4}(-7 \cos \alpha k_y  +2 m_1 \sin\beta k_y),\\ 
&-\cos k_x, \sin k_x, -\frac{1}{2}\cos k_x, -\frac{1}{2}\sin k_x,\\
& \frac{1}{4\sqrt{3}}(3 \cos\alpha k_y + 2(m_1 + 2 m_3) \sin\beta k_y)\big].
\end{split}
\end{align}

\subsection{Setup} \label{model_build}
Next we discuss how to obtain such a Hamiltonian with realistic crystal structure and orbital symmetry. The phase dependent off-diagonal term $e^{\pm ik_x}$, and the Zeeman term $\sin(\beta k_y)\sigma_z$ term can be simultaneously obtained in a tripartite lattice by applying linearly polarized light on each sublattices. At the end of this section, we discuss how to design such a lattice.

\subsubsection{Tight-binding (TB) model for electron-photon coupling induced $\sin{(\beta k_y)}$ term}\label{Sec:EPcoupling}
The motivation for the origin of $\sin{\beta k_y}$ term can be drawn from the fact that the dipole interaction between an electron with momentum ${\bf p}=\hbar{\bf k}$ and an EM wave with potential ${\bf A}=A{\hat \epsilon}$ (${\hat \epsilon}$ is the light polarization) is $H_{\rm int}=-\frac{e}{m}{\bf p}\cdot{\bf A}=-\frac{e\hbar A}{m}{\bf k}\cdot{\hat \epsilon}$. We choose a linearly polarized light with its polarization oriented along the $y$-direction. We take a single electron Hamiltonian under the periodic potential $U({\bf r})$ of the lattice as $H =\frac{ p^2}{2m^*} + U({\bf r})$. The corresponding Bloch wavefunction is $\eta_{\bf k}=\frac{1}{\sqrt{N}}\sum_{n}e^{i{\bf k}\cdot{\bf R}_n}u_n({\bf r})$, where $N$ is the total number of unit cells, $u_n({\bf r)}$ is the Wannier state at the $n^{\rm th}$ site located at ${\bf R}_n$. In the presence of vector potential ${\bf A}$, the Hamiltonian becomes $H^{\prime} =\frac{ ({\bf p}-e{\bf A})^2}{2m^*} + U({\bf r})$. For the EM wave, the spatial dependence of ${\bf A}$ can be neglected, and thus, the translational symmetry of the lattice remains the same. Therefore, the new Bloch wavecfunction simply changes to $\eta^{\prime}_{\bf k}=\frac{1}{\sqrt{N}}\sum_{n}e^{i{\bf k}\cdot{\bf R}_n}u^{\prime}_n({\bf r})$, where $u^{\prime}_n({\bf r)}=u_n({\bf r})e^{i\frac{e}{\hbar}\int_{{\bf R}_n}^{\bf r}{\bf A}\cdot d{\bf l}}=u_n({\bf r})e^{i\phi_n({\bf r})}$. $\phi_n({\bf r})$ is called the Peierls phase at ${\bf r}$ acquired by the charged particle in traversing from the $n^{\rm th}$ lattice site. It can be shown that $H^{\prime}|u^{\prime}_n({\bf r)}\rangle =e^{i\phi_n({\bf r})}H|u_n({\bf r)}\rangle$. Using these ingredients, we can now derive the tight-binding dispersion as 
\beq\label{TBdispersion}
\xi({\bf k}) &=& \langle \eta^{\prime}_{\bf k}| H^{\prime} | \eta^{\prime}_{\bf k}\rangle\nonumber\\
&=& \frac{1}{N}\sum_{n,n^{\prime}} e^{i{\bf k}\cdot({\bf R}_n-{\bf R}_{n^{\prime}})}  \int d{\bf r} \langle {u^{\prime}_{n^{\prime}}}| H^{\prime}| u^{\prime}_n \rangle\nonumber\\
&=& \frac{1}{N}\sum_{n,n^{\prime}} e^{i{\bf k}\cdot({\bf R}_n-{\bf R}_{n^{\prime}})}  e^{i (\phi_n-\phi_{n^{\prime}})}  \int d{\bf r} \langle {u_{n^{\prime}}}| H| u_n \rangle\nonumber\\
&=& \sum_{n,n^{\prime}} t_{nn^{\prime}}e^{i{\bf k}\cdot({\bf R}_n-{\bf R}_{n^{\prime}})}  e^{i (\phi_n-\phi_{n^{\prime}})}.
\eeq
Here $t_{nn^{\prime}}=\frac{1}{N}\int d{\bf r} \langle {u_{n^{\prime}}}| H| u_n \rangle$ is the TB hopping amplitude between $n$ and $n^{\prime}$ sites {\it without} the vector potential. We here restrict ourselves to the nearest neighbor hopping, i.e., $n^{\prime}=n\pm 1$. Let the lattice constant along the $y$-direction be $b$. By setting $t_{n(n\pm1)}=t_y$, and $\pm\phi=\phi_n-\phi_{n\pm1} =\pm\frac{e}{\hbar}Ab= \frac{e}{\hbar}Ab{\hat y}\cdot{\hat \epsilon}$, we obtain,
\beq\label{TBdispersion2}
\varepsilon({\bf k}) &=& t_y\left[e^{i(k_yb + \phi)}  + e^{-i(k_yb + \phi)}\right].\nonumber\\
%&=& 2t_y\cos{(k_yb + \phi)}.\nonumber\\
&=& 2t_y\left[\cos{(k_yb)}\cos{\phi} -\sin{(k_yb)}\sin{\phi}\right].
\eeq
We absorb $\cos{\phi}$ in to the TB term as $t_y(\phi)=2t_y(0)\cos{\phi}$, and define $m(\phi)=-2t_y\sin\phi$. Then we see that Eq.~\eqref{TBdispersion2} is the same as Eq.~\eqref{diagonal}. From this definition, it is easy to see that as the direction of polarization is reversed, $\phi\rightarrow -\phi$, and thus $m\rightarrow -m$, $t_y\rightarrow t_y$ while the perpendicular polarization yields $m (\phi=0) = 0$, and $t_y$ remains the same.  

\subsubsection{Tripartite lattice}\label{Sec:lattice}

For the SU(2) case, the phase dependent hopping term $\sim t\exp(ik_x)$ is obtained in bipartite lattice (c.f. Su-Schrieffer-Heeger model in 1D,\cite{ssh_rmp} or honeycomb lattice\cite{castroneto} in 2D) or for hopping between even and odd-parity orbitals.\cite{WeylTD} Similarly, for the SU(3) case, we need the same term for all three off-diagonal terms. Therefore, we propose a tripartite lattice as depicted in Fig.~\ref{model}. Also note that, the complex hopping term only includes $k_x$ terms, implying that different basis elements should be aligned along the $x$-direction only. Therefore, we consider three chains of different species which are connected via quantum tunneling in both directions. We assume periodic boundary conditions along both directions. 

The nearest neighbor hoppings along the $x-$direction between basis $1\rightarrow 2$, and $2\rightarrow 3$ give the same momentum dependence $e^{ik_x}$, while that for $1\rightarrow 3$ is $e^{-ik_x}$ (we set the corresponding hopping amplitudes as $t^{12}_x = -t_x, t^{13}_x = -t_x$ and $t^{23}_x = -\frac{1}{2}t_x$).  This reversal of Bloch phase serves the purpose of chirality inversion along this direction. Caution should be taken when the next-nearest hopping term becomes turned on, whose $k$-dependence is given by $-2t_{2y}\cos{k_y}e^{\pm ik_x}$, respectively. Such term adiabatically destroys the integer Chern number. Therefore, to avoid it we propose to increase the inter-atomic distance between adjacent chains to be as large as possible so that the hopping term $t_{2y}\rightarrow 0$. This setup simultaneously produces a $\cos{k_x}$ term for the intra-basis dispersions in Eq.~\eqref{diagonal} due to nearest neighbor hopping.

\subsubsection{Design principles}\label{Sec:design}

\begin{figure}[t]
%\hspace{-2.2cm}
%\centering
\includegraphics[width=\columnwidth]{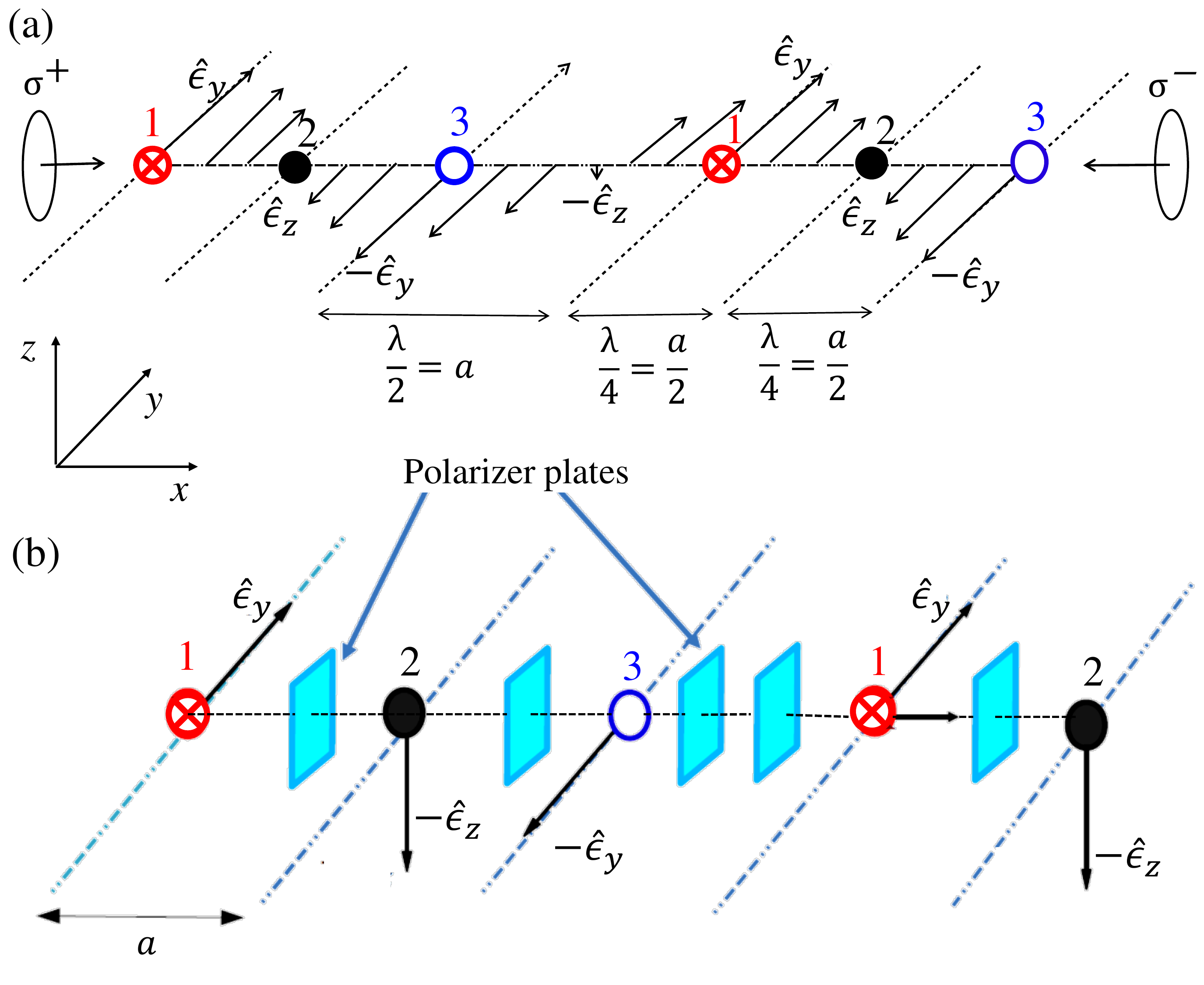}
\caption{(Color online) Schematic diagram of a possible experimental set-ups to realize a site-selective polarization. (a) Sisyphus cooling technique gives polarization gradient in a 1D lattice, when counter-propagating circularly polarized waves $\sigma^{\pm}$ are used. This creates a linear polarization that rotates in space (at $x=0$ polarization is along $\hat{\epsilon}_y$, at $x=\lambda/4$ and $\lambda/2$ polarization along $\hat{\epsilon}_z$ and $-\hat{\epsilon}_y$, respectively). Atoms are trapped at these $x$-values to create the necessary odd parity Zeeman term.
b) A linearly polarized EM wave along $\hat{\epsilon}_y$ is incident on the atom trapped in the 1st atomic chain. A polarizer plate, placed between the 1st and 2nd wires, rotates the polarization vector by $\pi/2$. The EM wave in the 2nd wire is then polarized along $\hat{\epsilon}_z$. Another polarizer plate, between the wires with the 2nd and the 3rd atom, rotates the incident polarization vector $\hat{\epsilon}_z$ by $\pi/2$ again, and the polarization vector along the 3rd wire is along $-\hat{\epsilon}_y$. Two quarter-wave polarizer or one half-wave polarizer plates are need to be placed between the 3rd and 4th wire (site index 1 as periodicity is imposed) to rotate the polarization vector from  $-\hat{\epsilon}_y$ to  $\hat{\epsilon}_y$. This generates the desired polarization gradient in our lattice grid structure.}
\label{plate_diag}
\end{figure}

For all the above terms, no constraint arose about the specific parity (or orbital symmetry) of each basis. Therefore, coupled chain structure can be engineered with ultracold fermionic or bosonic atoms in optical lattice setup, or with quasi-1D quantum wires of electrons with lithography or pulse laser deposition method. However, for the generation of site-selective $\sin{\beta k_y}$ term with light-matter interaction, specific structure or tuning is required. We suggest two practical experimental setups for the engineering of this phenomenon. Of course, the possibilities are abundance.

{\it (a) Sisyphus Cooling:-} Sisyhus cooling technique provides a spatial modulation of the static polarization.~\cite{sisyphus1,sisyphus2} In this technique, two counter-propagating laser beams with orthogonal polarization are used to create standing wave with a polarization gradient that alternates between circular and/or linear polarization with right and left handedness. Atoms trapped by the laser can acquire different polarization according to their positions with respect to the wavelength of the lasers. By tuning the laser wavelength with respect to lattice constant of the tripartite lattice, we can construct the desired site-selective polarization as follows.   

In Fig.~\ref{plate_diag}(a), we demonstrate the setup. We take two counter-propagating, oppositely oriented circularly polarized ($\sigma^{\pm}$) EM fields $(\hat{\epsilon}_y+i\hat{\epsilon}_z)e^{i {\bf q}.{\bf x}}$ and $(\hat{\epsilon}_y-i\hat{\epsilon}_z)e^{-i {\bf q}.{\bf x}}$ to trap atoms along the $x$-direction. $q=2\pi/\lambda$ is the light's wavevector and $\lambda$ is its corresponding wavelength (the frequency dependence of the field do not make any contribution to our analysis and thus not discussed). The resultant field $2[\hat{\epsilon}_y\cos{(q x)}-\hat{\epsilon}_z \sin{(qx)}]$ has a spatially dependent polarization. At $x=0$, it starts off with a linear polarization along the $\hat{\epsilon}_y$-direction, at $x=\lambda/4$ the polarization is along the $\hat{\epsilon}_z$-direction, and at $x=\lambda/2$ it is rotated along the $-\hat{\epsilon}_y$-direction. In Fig.~\ref{model}, we assume sites 1 and 3 have opposite polarization along the $y$-direction, while site 2 has orthogonal polarization. Therefore, to implement the Sisyphus technique, site 1 sits at $x=0$, while site 3 resides at $x=a=\lambda/2$. Then the distance of site 2 from 1 and 3 is $a/2$ to gain the orthogonal polarization ($\hat{\epsilon}_z$). Since the system is confined in 2D, it has no momentum along the $z$-direction and thus no dipole term arises for site 2. With this atomic position, the corresponding Bloch phases for hopping $1\rightarrow 2$ and $2\rightarrow 3$ is $e^{ik_x/2}$, while that for $1\rightarrow 3$, and $e^{-i k_x}$. This gives the Chern number (see calculations in Sec.~\ref{cal_chern} as $(-2,4,-2)$.

Discussions of the advantage and limitation of using this technique are in order. In this specific Sisyphus technique, the resulting EM field has only linear polarization along the chains where the atoms are placed \cite{sisyphus1}. In the ground state, atoms have hyperfine levels $g_s=\pm 1/2$. For this two states, the light shift caused by the interaction between the atoms and EM field is exactly equal, and also it does not vary with the direction of propagation (here $x$-direction). This is a great advantage to our setup where we do not have to deal with the light-shift splitting of the hyperfine levels coming from the interaction with the EM field. However, this technique has a limitation. The lowest temperature that can be attained by this cooling is set by the recoil energy $\hbar^2 q^2/2m$, which is the kinetic energy an atom gains after absorbing a photon. In most alkali atoms, this temperature is below $1\mu K$. At such low temperatures, the atomic de Broglie wavelength becomes comparable to the cooling laser wavelength (though still shorter than is required for the BEC phase transition) and hence to the extent of the potential wells. It is therefore no longer possible to localize the atomic wave packet in the potential wells, even if they were deeper than the photon recoil energy. However, this limitation has been successfully overcome, and there are numerous usage of this technique in the literature (see e.g. Refs.~\cite{sisyphus3,sisyphus4}).

{\it (b) Using polarizer plates:} Another simpler setup can be made using polarizer plates between consecutive atomic chains, which rotates the polarization of the incident em field by $\pi/2$, as illustrated in Fig.~\ref{plate_diag}(b). This can be repeated periodically along the $x$-direction (propagation direction of the incident EM field) to achieve the desired polarization reversal along each chain. We start with a linearly polarized light ($\hat{\epsilon}_y$), incident on site 1.   
A quarter-wave polarizer between site 1 and 2, rotates the polarization to $\hat{\epsilon}_z$, having no dipole interaction here. Again another quarter-wave polarizer between site 2 and 3, rotates the polarization along the $-\hat{\epsilon}_y$. To achieve periodic boundary conditions, the next site 4 (which should be equal to site 1), we need the polarization to be along $\hat{\epsilon}_y$, which can be achieved by either 2 quarter-wave or one half-wave polarizer. This configuration is repeated along the x-direction. Polarizer plates, like half-wave and quarter-wave plates, are associated with intensity losses, which can be encountered by replacing them with birefringent filter plates. Polarizer plates may attenuate the hopping amplitudes between the atoms. However, since the value of the Chern number does not depend on the hopping amplitudes, the system will remain topologically invariant as long as the tunneling is finite. 

{\it (c) Other possible condensed matter setups:} Few other possible engineering principles can be envisioned using condensed matter setups. A site-selective electric field grid can be created in 1D lattices. (i) Alternatively, it is shown recently that high-energy photons can be used to selectively photonize the valence electrons of hydrogen chloride by using the rotational dependence of the photonization profiles.\cite{siteselective_photoemission} (ii) One can also pursue a possibility of using ferroelectric substrate in which due to in-plane inversion symmetry breaking, electric field polarization in different layer or chain can induce site-selective polarization via proximity effect to the top lattice of our interests. (iii) We can also use orbital-selective chains. Recalling that the $\sin{k_y}$ term is absent in the second diagonal term (Eq.~\eqref{diagonal}), we can think of an orbital symmetry for the second basis which is orthogonal to the direction of light's polarization. Corresponding choices of orbitals are $p_x$- or $d_{xz}$ orbitals for which $H_{\rm int}=0$ with polarization along the $y$-direction. For the other two orbitals, we can consider combinations such as $s$- and $p_y$ orbitals, etc. 

\section{Band topology and Berry curvature}\label{cal_chern}

\begin{figure}[t]
%\hspace{-2.2cm}
\centering
\includegraphics[width=\columnwidth]{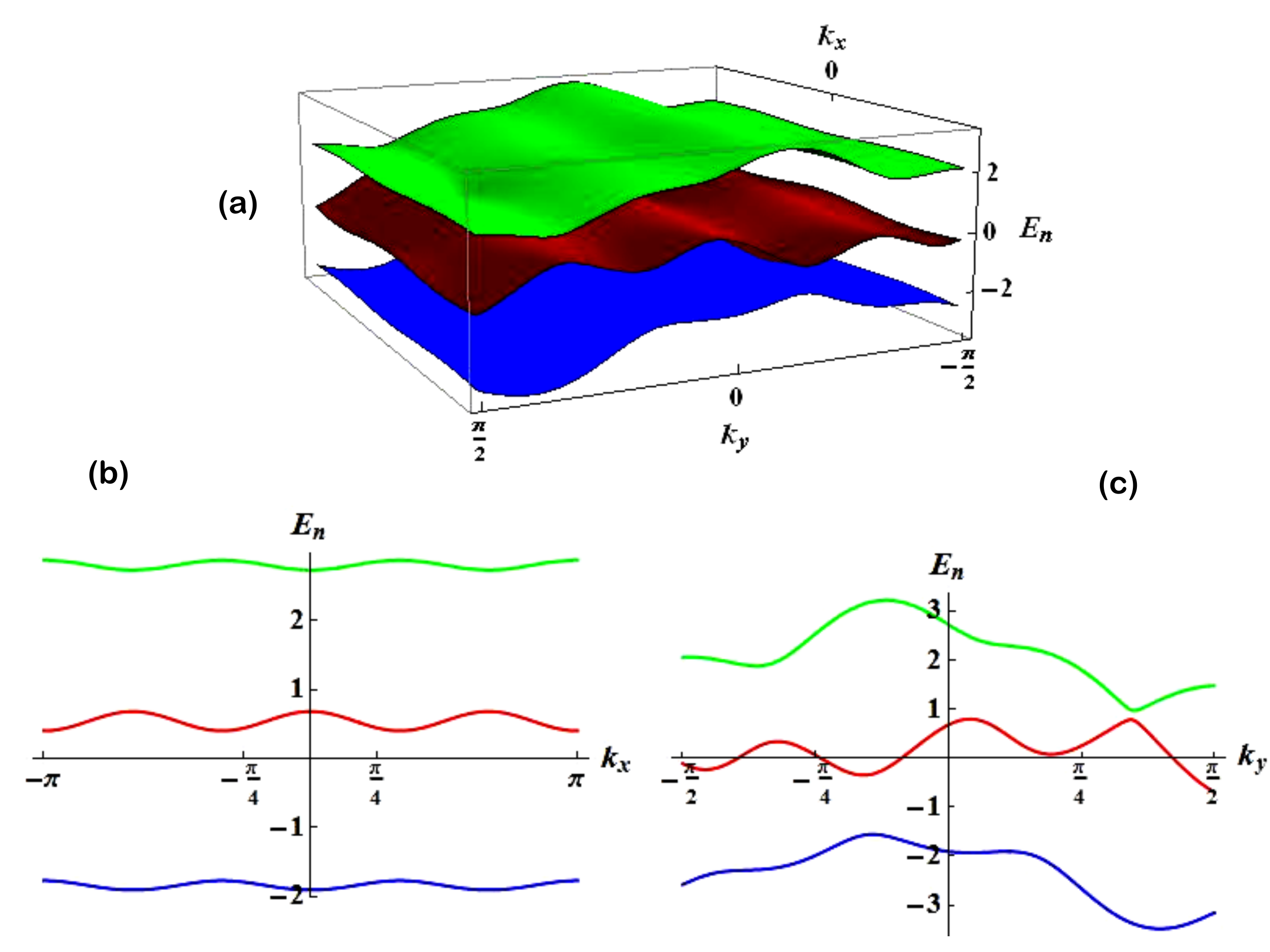}
\caption{(Color online) Bands demonstrating the energy dispersion for the proposed model are shown. (a) Surface plot with both $k_x$ ($-\pi$ to $\pi$) and $k_y$($-\frac{\pi}{2}$ to $\frac{\pi}{2}$), (b) Keeping $k_y=0$, dispersion plot along $k_x$, from $-\pi$ to $\pi$,  and (c)  Keeping $k_x=0$, dispersion plot along $k_y$, from $-\frac{\pi}{2}$ to $\frac{\pi}{2}$. All quantities are measured in units $t_x=t_y=1$, and $|m_1|=|m_3|=\sqrt{3} t_y$. 
 }
\label{3Dbren}
\end{figure}

\begin{figure*}[t]
%\hspace{-2.2cm}
%\centering
\rotatebox[origin=c]{0}{\includegraphics[width=1.8\columnwidth]{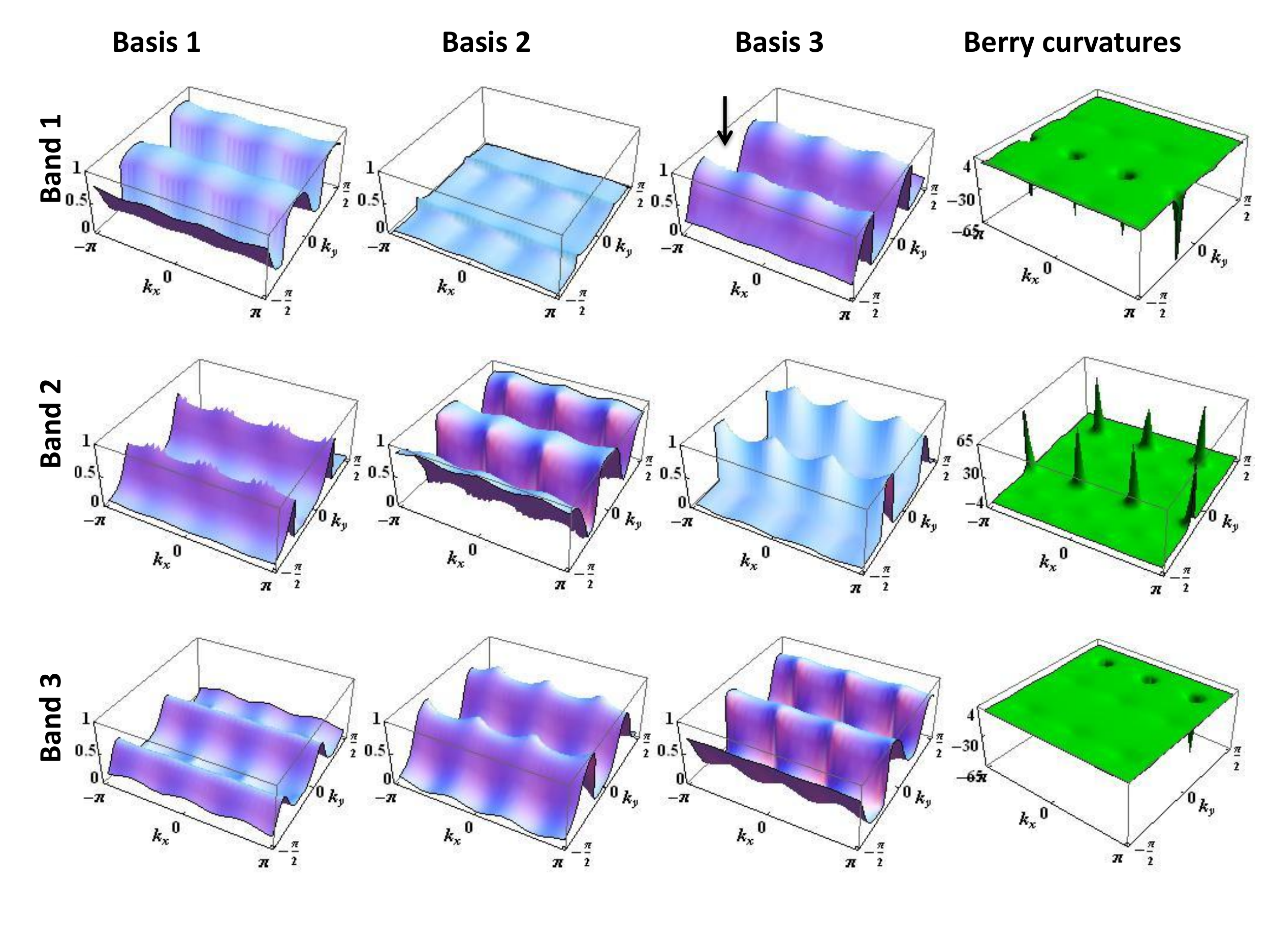}}
%\vspace{12pt}\\
\caption{(Color online) First three columns: The orbital weights of the basis states $\psi_j$ ($j=1,2,3$) are shown for each energy bands $E_1$, $E_2$ and $E_3$ respectively. The arrow in the figure corresponding to the third column shows the position of one minima which is corresponding to the gapless point of one of the edge states in the system. Fourth column: The berry curvature $\Omega_i (i=1,2,3)$ corresponding to each energy band is shown in the fourth column. The berry curvatures show a sharp peak at the corresponding gapless points of the edge states, indicative of a band inversion and a topological phase transition at the respective $k$-points. }
\label{supp_band}
\end{figure*}

In Fig.~\ref{3Dbren}, we plot the band structure in the momentum-space for the parameter values of $m_1=-m_3=\sqrt{3}t$ and $\alpha=\beta=2$, and $t_x=t_y$. The electronic structure consists of three well separated bands, with only Dirac-like nodes at various discrete non-high symmetric $k$-points (see Fig.~\ref{3Dbren}(a)). Therefore, a topological invariant can be separately assigned for each band. However, projecting the orbital character onto each band, we observe that substantial exchange of orbital character occurs in each band. We visualize the three orbital characters (in different row) for three different bands (in different column) in the entire 2D $k$-space in Fig.~\ref{supp_band}. 

As discussed in the introduction section, band inversion is an important criterion for both the SU(2) and SU(3) topological classes. In time-reversal invariant SU(2) topological classes, bands are only required to be inverted at the time-reversal invariant $k$-points. This makes it easier to define the band inversion strength simply by defining the band gap between the two bands at the time-reversal invariant $k$-points.\cite{AdiabaticTI,TIreviewTD} Such simple definition becomes difficult to implement for SU(3) materials. On the other hand, we recognize that the Berry curvature acquires spike at the discrete band degenerate $k$-points ($k$-space monopoles) across which bands are inverted in two orthogonal directions (the two orthogonal directions do not necessarily have to be along the $k_x$- and $k_y$-directions). In this spirit we can define a band inversion strength via occupation number or the `orbital weight' (here orbital refers to the SU(3) components) of the band at each $k$-points. Another interesting feature of the monopole is that it represents a saddle point in the orbital weight (see Fig.~\ref{supp_band}), in the sense that if an orbital character obtains a maximum along the $k_x$ direction, it obtains a minimum in the $k_y$ direction. The rightmost column of Fig.~\ref{supp_band} refers to the $k$-resolved Berry curvature. We see that at all the ${\bf k}^*$-points where $\Omega({\bf k}^*)$ diverges in a given band, the corresponding orbital weight profile exhibits a saddle point.  

Let us define a quantity $\kappa^{\nu}_{k_{l}}={\rm sgn}\left[ |\gamma^{\nu}_i(k_l)|^2-|\gamma^{\nu}_{j\ne i}(k_l)|^2\right]$, where $\gamma^{\nu}_{i}$ is the eigenfunction of the Hamiltonian in Eq.~\eqref{Ham1} corresponding to $\nu^{\rm th}$ band and $i^{\rm th}$ SU(3) spinor component. In the $\nu^{\rm th}$band, if the $i^{\rm th}$ orbital obtains a maximum (and $j^{\rm th}$ orbital then obtains a minima) along, say $k_x$-direction, then $\kappa^{\nu}_{k_{x}}=+1$. On the other hand, its minimum would corresponds to $\kappa^{\nu}_{k_{x}}=-1$. Therefore, the locus of the monopole is defined by the discrete points which satisfy $\kappa^{\nu}_{k_{x}}\kappa^{\nu}_{k_{y}}=-1$. This is of course an indication of the location of finite Berry curvature. A singularity in $\Omega(k)$ is obtained where the bands are fully degenerate. 

For the higher energy band, among the six visible saddle-points, three of them reside in the $-k_y$ region. They give three negative spikes in the Berry curvatures, as shown in the corresponding rightmost column of Fig.~2. (The two extreme peaks occurring at the zone boundary are related by reciprocal lattice vectors). This gives the corresponding Chern number $C_n=\sum_{\bf k} \Omega_n({\bf k})$ (where $n$ stands for bands) to be -3. In this band, inversion occurs between orbitals 1 and 3  across the three saddle points. In the lowest band, the Berry curvature peaks occur in the corresponding $+k_y$ side due to the band inversions between orbitals 2 and 3. The Chern number of this band comes out to be the same as -3. The middle band shows band inversions between all three orbitals at the same locations in both $\pm k_y$ sides, with positive Berry curvatures and thus obtains a Chern number of $+6$. 

It can also be shown that the Chern number does not depend if the strength of the Pierels' phase is different in the different orbitals. As long as the $\sin{\beta k_y}$ terms have opposite signs in the diagonal terms (i.e. mimicking a Zeeman-like term), the model is topologically non-trivial. Changing the off-diagonal elements, say from $e^{i k_x}$ to $e^{2 i k_x}$, we find higher  Chern number $(-4,8,-4)$. In the Appendix below, we discuss several other parameter sets where Chern number can be tuned by different values of $\alpha$, $\beta$, and other terms in the Hamiltonian.

\section{\label{cal_edge} Calculation of edge states} 

\begin{figure}
%\hspace{-2.2cm}
\centering
\includegraphics[width=1.0\columnwidth]{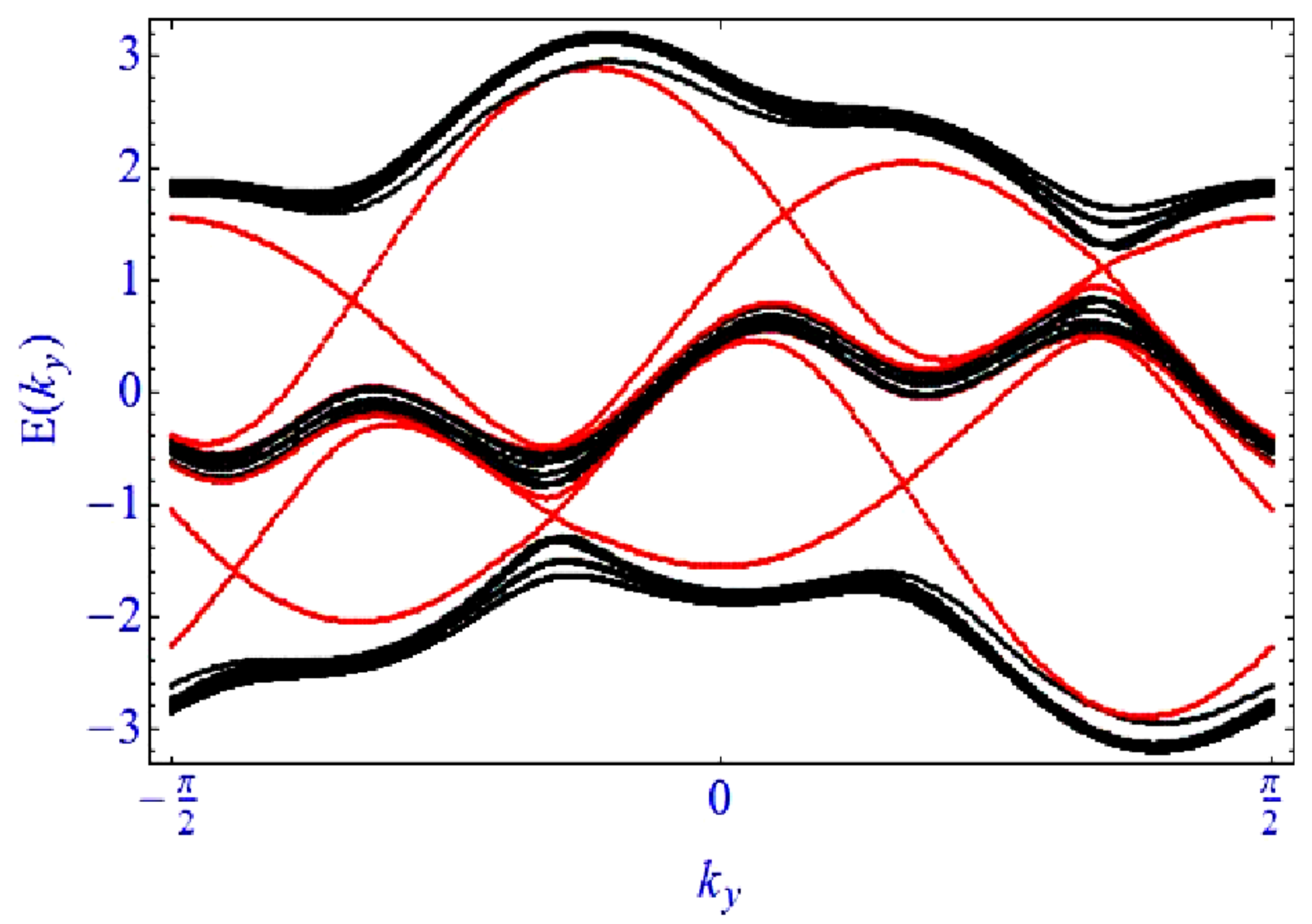}
\caption{(Color online) Edge state structure for $N=30$, with $\alpha=\beta=2$ and $|m_1|=|m_3|=\sqrt{3} t_y$. Also, $t_x=t_y=1$. There are six gapless points within the first Brillouin zone, as indicated by the chern numbers being $\{-3,6,-3 \}$.
}
\label{edgestate}
\end{figure} 

Non-trivial topological character can be observed from the edge state dispersion and non-local electrical measurements.\cite{TIreviewTD,TIreviewCK,TIreviewSCZ} We study the characteristics of the edge state parallel to the $y$-direction for the above parameter set which gives Chern numbers $(-3,6,-3)$. Therefore, we solve the Hamiltonian in Eq. ~\ref{Ham1} with periodic boundary conditions in the $y$-direction, and open boundary condition in the $x$-direction with a finite size lattice of $N$ atoms. Considering $\Phi_i({\bf k}_y)$ as the Wannier state localized on the $i^{\rm th}$ atom we can now expand the Hamiltonian in a $N\times N$ matrix as
\begin{eqnarray}\label{edge}
H(k)&=& -\sum_i\Big[\Phi_i^\dagger(k_y) {\mathcal{B}_{i,i}}(k_y)\Phi_i(k_y)\nonumber\\
&&\quad\quad+ \Phi_i^\dagger(k_y) {\mathcal{A}_{i,i+1}}\Phi_{i+1}(k_y)\Big] + {\rm h. c.}, 
%+\Phi_{i+1}^\dagger(k_y) {\mathcal{A}}^\dagger\Phi_{i}(k_y)]\,.
\end{eqnarray}
\begin{align} \label{bmatrix}
\begin{split}
{\rm where,}~~
{\mathcal{B}}(k_y)=&\cos {\alpha k_y} \begin{pmatrix} 
-2t_y&0&0\\
0&\frac{3}{2}t_y&0\\
0&0&-t_y\\
\end{pmatrix}\\
+
\sin {\beta k_y}
&\begin{pmatrix}
-\frac{2m_1-m_3}{3}&0&0\\
0&\frac{m_1+m_3}{3}&0\\
0&0&\frac{m_1-2m_3}{3}\\
\end{pmatrix}\,,
\end{split}
\end{align}
and,
\begin{eqnarray}\label{amatrix}
{\mathcal{A}}= t_x\begin{pmatrix}
0&0&-1\\
-1&0&0\\
0&-1/2&0\\
\end{pmatrix}\,.
\end{eqnarray}
Eigenvalues of the Eq.~\eqref{edge} are plotted in Fig.~\ref{edgestate} with the same parameter set. The essence of the topological edge state is that it must adiabatically connect to the bulk states which is clearly observed in the present case. Owing to the maximum Chern number of 6, there are 6 edge states also (red lines). Edge states with opposite dispersion, connecting to different bulk bands with opposite sign of the Chern number, meet at discrete ${\bf k}^*$-points where the Berry curvature obtained singularities in Fig.~\ref{supp_band}. Consistently, there are total of 6 such band touching points for the edge states.

We see that $\mathcal{B}$ term is diagonal in this basis which gives the dispersion along the $k_y$ direction. These states become gapped by $\mathcal{A}$. However as the number of lattice site is increased, the gap at the edge state vanishes at the $k_y$-points where Berry phase acquires divergence. Therefore, expanding the edge state near these points, we find that three eigenstates up to linear-in-$k_y$ as (substituting $|m_1|=|m_3|=m$)
\begin{eqnarray} \nonumber
E_1(k_y) &=& -t_y  \bigg(2 + \frac{m}{t_y} k_y \bigg) \nonumber \\
E_2(k_y) &=& \frac{3}{2}t_y \nonumber \\
E_3(k_y) &=& -t_y \bigg(1 - \frac{m}{t_y} k_y \bigg).
\end{eqnarray}
We notice that the second term represents a localized bound state (soliton), while the other two bands are linear with $k_y$ in the low-energy region. As $\mathcal{A}$ term is turned on, these three states spit into six states, in accordance with the higher Chern number in the bulk state.

\section{\label{spin} Extension to spinful case}

In our model, SU(3) symmetry is obtained for spinless fermions, in which spin of the particles is a dummy variable. This Hamiltonian respects the spin-rotational symmetry. Therefore, as long as this symmetry is held (in the absence of spin orbit coupling or magnetic moment), the topological invariant remains the same for all values of spin in a given system, that means for both fermionic and bosonic systems. We now discuss how the result changes when the spin rotational symmetry is broken. For generality, we assume the atoms/electrons have a spin value $S$, which splits into 2$S$+1 multiplets once the rotational symmetry is lifted. In this case, our starting SU(3) spinor has the dimension of 3(2$S$+1). For illustration we take $S=1/2$, while the obtained conclusions below remain the same for any other values of $S$. In this case, the spinor is $\Psi_k^{\dag}=\left( {\psi}_{1\uparrow}^{\dag}(k), {\psi}^{\dag}_{2\uparrow}(k),{\psi}^{\dag}_{3\uparrow}(k), {\psi}_{1\downarrow}^{\dag}(k), {\psi}^{\dag}_{2\downarrow}(k),{\psi}^{\dag}_{3\downarrow}(k)\right)^{T}$, in which the Hamiltonian becomes a $6\times 6$ matrix. The Hamiltonian can be split into two $3\times 3$ diagonal blocks as $H_{\uparrow\uparrow}$, and $H_{\downarrow\downarrow}$, and off diagonal blocks $H_{\uparrow\downarrow}$, and $H_{\uparrow\downarrow}^{\dagger}$. In the absence of the spin-flip terms, i.e., when $H_{\uparrow\downarrow}=0$ at all momentum, the Hamiltonian breaks into a block diagonal one. As the time-reversal symmetry is broken by introducing an exchange term $E_z$, the block diagonal terms become $H_{\uparrow\uparrow/\downarrow\downarrow}=H\pm E_z I_{3\times 3}$, where $H$ is given in Eq.~\eqref{Ham1} above. Note that Chern number does not depend on an overall energy shift ($a(k)$ term in Eq.~\ref{su3_gellmann}). Therefore, the system remains topologically non-trivial. If the exchange energy $E_z$ is increased beyond the band gap (determined by $b_3$ and $b_8$ terms), it can introduce new band inversion. Depending on the nature of the band inversion, the system can obtain either higher or lower Chern number (a topological phase transition). 

In what follows, as long as a simple exchange field is present to split the spin states, such as magnetic field or Ising like ferromagnetic state, the system maintains its topological property. We have also checked that when the exchange energy is taken to be orbital dependent (breaking parity), the system continues to obtain the same Chern number untill a new band inversion occurs. When the spin-flip term $H_{\uparrow\downarrow}$ is introduced for the case of spin-orbit coupling or antiferromagnetism, spin is no longer a good quantum number. In this case, Chern number cannot be defined for each spin state of band. Therefore, there will be a topological phase transition into either a trivial case, or to another topological class, such as $Z_2$ family for fermions. We discuss one such possibility using Ising spin-orbit coupling. 

Recently, there is a number of transitional metal dichalcogenide materials synthesized in isolated monolayers which have buckled honeycomb lattice structure. The buckled structure looses inversion symmetry in all three directions. This causes a large out-of-plane spin polarization, similar to Ising spin which then becomes locked to the only one momentum component via SOC. The corresponding SOC component, as known by Ising SOC, is expressed by $H_{ISOC} = \alpha_I d_z({\bf k})\sigma_z$, where $\alpha_I$ is the SOC strength and $z$-axis is chosen as the easy axis of the Ising spins. For a buckled honeycomb lattice, it is shown that the momentum dependent term comes out to be\cite{IsingSOC} $d_z({\bf k}) = \sin{( k_y)} - 2\cos{(\sqrt{3}/2k_x)}\sin{( k_y/2)}$. This term essentially gives an odd-parity (along $k_y$) Zeeman-like term. To go from a SU(2) system to SU(3), one requires another spinless (or singlet) U(1) basis. One possible way to obtain such a Hamiltonian is to introduce a spin-singlet Kondo impurity as new basis in each unit cell. Solving a Kondo lattice model, one can obtain the reqired Bloch phase $e^{ik_x}$ for the hopping between SU(2) species to the Kondo impurity. Finally, the Rashba-SOC between the SU(2) basis provides the chirality between them. In this way, an effective SU(3) topological Hamiltonian can be constructed which can be written in terms of $\hat{\lambda}$-matrices. The corresponding results will be published elsewhere \cite{Isingpaper}. \hphantom{} \\

\section{ \label{conclusion} Discussion and Conclusion}

The paper delivers the following messages. We showed that for the SU(3) topological phase, two of the basis components must contain counter-propagating chiral phase. The third component can take any of the chirality and/or may be chiral-free. This situation is analogous to a SU(2) topological insulator in which chirality inversion is also an important criterion. One way to obtain chirality inversion in SU(2) systems is through reversing the SOC, either intrinsically in layered materials,\cite{TIreviewTD} or in engineered heterostructures in which adjacent layers are assumed to have opposite SOC,\cite{EngineeringTI,RayOpticalLattice} or even through Fermi surface nesting between spin-orbit coupled objects which in special cases can render chirality reversal.\cite{Gaurav}  

We engineered the complex phase dependent off-diagonal terms in a tripartite lattice through uncompensated Bloch phase. The chirality inversion is naturally obtained between the sublattices 1 \& 3, which is reversed from that between 1 \& 2 and 2 \& 3. The multi-channel set-up of one-dimensional atomic chains suggested in our model can be visualized as an array of quantum wires, hosting different types of orbitals, being subjected to a linearly polarized static electromagnetic field. Quantum wires are being studied extensively to describe varied topological phenomena theoretically~\cite{haim,seroussi,sela,alicea}. Fractional topological phases are being studied in weakly coupled quantum wires, in both two and three dimensions.\cite{lubensky,mudry,teo,yoreg} Periodically driven systems can also show interesting non-trivial topological effects in spinless systems ~\cite{yan,fulga,ezhao,diehl,goldman} and can also be extended to SU(3) systems. Furthermore, Fermi surface nesting in quantum wires with SOC can also lead to chirality inversion between different sublattices through the formation of an exotic spin-orbit density wave, as predicted \cite{SODWth} and subsequently realized \cite{SODWexp} in Pb-, and Bi-based quantum wires. Quantum wires can be grown on electric field grid or on ferroelectric subtracts which can be tailored to obtain chirality inversion in the off-diagonal and diagonal terms. 

The second criterion for the SU(3) topological Hamiltonian is unique to this system. Here, two of the diagonal terms in the Hamiltonian must contain an odd parity term, such as a sinusoidal function of the momentum. This poses an important bottleneck to engineer SU(3) topological phase in a condensed matter setup. Here we mainly focused on designing such term through electromagnetic wave by generalizing the tight-binding Hamiltonian with an applied vector potential. This gives a spinless SU(3) topological material. We proposed two realistic techniques, using either the Sisyphus cooling technique or polarizer plates to obtain the desired site-selective polarization. 

Robustness of the spinless SU(3) topological phase when spin is introduced is also discussed. We showed that as long as there is only a magnetic exchange term present  without any spin-flip term, the topological invariant is robust upto a new band inversion. As the exchange energy surpass the band gap, a topological phase transition occurs, reducing or increasing the Chern number by an integer value. When a spin-flip term is introduced (such as spin orbit coupling or antiferromagnetic like term), spin is no longer a good quantum number, and Chern number can no longer be defined for each spin or band. So our formalism does not hold any more. 

The presence of edge states can be detected by using non-local electrical measurements~\cite{Roth,gusev}, where a current applied between two probes creates a net current along the edge. This current is then measured by a pair of voltage probes which is placed away from the bulk current path to avoid dissipation. \hphantom{} \\

\begin{acknowledgments} 
We acknowledge Kallol Sen for valuable discussions during the course of the work. AG acknowledges the financial support from Science and Engineering Research Board (SERB), Department of Science \& Technology (DST), Govt. of India for the National Post Doctoral Fellowship. TD acknowledges acknowledges the financial support from the same board under Start Up research Grant (Young Scientist). % We are thankful to DST and CSIR for funding this research.
\end{acknowledgments}

\appendix

\section{\label{otherhamiltonians} Other forms of Hamiltonians}

So far, we had considered a specific form of the most general Hamiltonian given in Eq.~\eqref{gensu3}. This model Eq.~\eqref{Ham1} is realized in a tripartite lattice with site-selective electromagnetic field. With respect to the structure of our model, the Hamiltonian can be rewritten in a more generalized form by expressing the off-diagonal terms as,
\beq
H_{12}=  t_x^{12}\ e^{-ik_x} \ ;  H_{13}=   t_x^{13} \ e^{-ik_x} \ ; H_{23}=   t_x^{23} \ e^{-ik_x} .
\label{Hamgen}
\eeq
Where the $t_{ij}$ provide the inter-basis hopping strengths between nearest neighbour sites. The diagonal terms are kept in the same form as in Eq. (\ref{diagonal}) with various choice of $t_i$ and $m_i$. Here we discuss various other combinations of the diagonal and off-diagonal terms which give finite Chern number, some of which may require different lattice structure than the triparticle lattice discussed in the main text. It should be noted that the following list is not necessarily exhaustive, and more combinations can be derived based on the basic principles deduced in the main text.  In all combinations, the Hamiltonian must respect the SU(3) symmetry and is represented by the eight Gell-Mann matrices. \hphantom{} \\

{\bf \underline{Case I: $\mathbf{\alpha=\beta=1}$}}
\begin{enumerate}
\item{ With $t_1 = t_3= -t_y$, $ t_2 = 2t_y, m_1 = -m_3 = -\sqrt{3}t_y$, $m_2=0$, $t_x^{12}=t_x^{13}=t_x^{23}=-t_x$, this Hamiltonian provides integer Chern number set $(-3,6,-3)$.
}
\item{Same as (1) but with $t_x^{12}=-t_x\cos k_y$. This Hamiltonian provides Chern numbers $(-1,2,-1)$.
}
\item{ With $t_2=m_2=0$ i.e. $\xi _2({\bf k})=0$, and $t_3=t_y/2 $ with rest of the coefficients as in 2. This Hamiltonian again, gives Chern numbers $(-3,6,-3)$.
}
\end{enumerate}

{\bf \underline{Case II: $\mathbf{\alpha=\beta=2}$}}\\
\begin{enumerate}
\item{With $t_2=m_2=0$ and $t_3=t_y/2 $, as in point 3 of Case I, replace $t_x^{12}=-t_x$, $t_x^{13}=2t_x \cos 2k_x$ and $t_x^{23}=-t_x \sin (k_y-\sqrt{3})$, keeping the rest of the coefficients same as in point 1 in Case I. This Hamiltonian gives the Chern numbers $(4,0,-4)$.
}
\item{The almost similar configuration as in point 1 in Case II, only changing $t_x^{12}$ as $t_x^{12}=\pm t_x(\cos k_y+\sqrt{3}\sin k_y)$ and using the same sign (either $+$ or $-$) for $t_x^{ij}$ i.e with $t_x^{13}=\pm t_x\cos k_x$ and $t_x^{23}=\pm t_x\sin(k_y-\sqrt{3})$ the Chern number for this hamiltonian is $(-3,0,-3)$.  
}
\end{enumerate}

In all the above calculations, $t_y=t_x=1$.


\begin{thebibliography}{1}
\bibitem{klitzing} K. V. Klitzing, G. Dorda, and M. Pepper,  Phys. Rev. Lett, \textbf{45},  494 (1980).

\bibitem{haldane} F. D. M. Haldane, Phys. Rev. Lett. {\bf 61}, 2015 (1988).

\bibitem{ssh_rmp} A. J. Heeger, S. Kivelson, J. R. Schrieffer, and W. -P. Su, Rev. Mod. Phys. {\bf 60}, 781 (1988).

\bibitem{TIreviewTD} A. Bansil, H. Lin, and T. Das, Rev. Mod. Phys. {\bf 88}, 021004 (2016).

\bibitem{TIreviewCK} M. Z. Hasan, and  C.L. Kane,  Rev. Mod. Phys.~{\bf 82}, 3045 (2010).

\bibitem{TIreviewSCZ} X. L. Qi, and  S.C.  Zhang, Rev. Mod. Phys.~{\bf 83}, 1057 (2011).

\bibitem{castroneto} A. H. Castro Neto, F. Guinea, N. M. R. Peres, K. S. Novoselov, and A. K. Geim, Rev. Mod. Phys. {\bf 81}, 109 (2009).

\bibitem{Balatsky}T. O. Wehling, A.M. Black-Schaffer, and A. V. Balatsky, Adv. Phys. {\bf 63}, 1 (2014).

\bibitem{Ando}Y. Ando, J. Phys. Soc. Jpn. {\bf 82}, 102001 (2013).

\bibitem{Vishwanath} O. Vafek, and A. Vishwanath, Ann. Rev. Cond. Mat. Phys. {bf 5}, 83 (2014).

\bibitem{WeylTD} T. Das, Phys. Rev. B {\bf 88}, 035444 (2013). 

\bibitem{EngineeringTI}T. Das, and A.V. Balatsky, Nat. Commun. {\bf 4}, 1972 (2013).

\bibitem{mukunda} G. Khanna, S. Mukhopadhyay, R. Simon, and N. Mukunda, Ann. Phys. (N. Y.)  {\bf 253}, 55 (1997).

\bibitem{georgi} H. Georgi, {\it Lie Algebras In Particle Physics: from Isospin To Unified Theories } (Benjamin/Cummings, Reading, MA, 1982) .

\bibitem{gellmann} M. Gell-mann and Y. Ne'eman, {\it The Eightfold Way} (Benjamin, 1964).

\bibitem{galitski} R. Barnett, G. R. Boyd, and V. Galitski, Phys. Rev. Lett, \textbf{109}, 235308 (2012).

\bibitem{jpsj} S. Y. Lee, J. H. Park, G. Go, and J. H. Han,  J. Phys. Soc. Jpn. {\bf 84}, 064005 (2015).

\bibitem{sisyphus1} J. Dalibard and C. Cohen-Tannoudji, J. Opt. Soc. Am. B {\bf 6}, 2023 (1989).

\bibitem{sisyphus2} D. Wineland, J. Dalibard and C. Cohen-Tannoudji, J. Opt. Soc. Am. B {\bf 9}, 32-42 (1992).

\bibitem{sisyphus3} P. Hamilton, G. Kim, T. Joshi, B. Mukherjee, D. Tiarks, and H. Müller, Phys. Rev. A {\bf 89}, 023409 (2014).

\bibitem{sisyphus4} H. J. Metcalf and P. van der Straten, J. Opt. Soc. Am. B {\bf 20}, 887–908, 2003.

\bibitem{siteselective_photoemission} C. Miron, Q. Miao, C. Nicolas, J. D. Bozek, W. Andraojc, M. Patanen, G. Simoes, O. Travnikova, H. Agren, and F. Gelamukhanov, Nat. Commun. {\bf 5}, 3816 (2014).

\bibitem{AdiabaticTI}H. Lin, T. Das, Y. J. Wang, L. A. Wray, S.-Y. Xu, M. Z. Hasan, A. Bansil, Phys. Rev. B {\bf 87}, 121202 (R) (2013). 

\bibitem{RayOpticalLattice}S. Ray, K. Sen, and T. Das, arXiv:1602.02926.

\bibitem{Gaurav}G. Gupta, and T. Das, Manuscript under preparation. 

\bibitem{haim} A. Haim,  A. Keselman,  E. Berg, and   Y. Oreg,  Phys. Rev. B {\bf 89}, 220504(R)  (2014).

\bibitem{seroussi} I. Seroussi,  E. Berg, and  Y. Oreg,  Phys. Rev. B {\bf 89}, 104523 (2014).

\bibitem{sela} Y. Oreg,  E. Sela,  and A. Stern,  Phys. Rev. B  {\bf 89}, 115402 (2014).

\bibitem{alicea} J. Alicea,	Y. Oreg, 	G. Refael,	F. von Oppen	,  and   M. P. A. Fisher, Nat. Phys. {\bf 	7},  412-417 (2011).

\bibitem{lubensky} C. L. Kane,  R.  Mukhopadhyay, and  T. C.  Lubensky,  Phys. Rev. Lett. {\bf 88} , 036401 (2002).

\bibitem{mudry} T. Neupert, C. Chamon, C. Mudry, and R. Thomale,  Phys. Rev. B {\bf 90}, 205101 (2014).

\bibitem{teo} J. C. Y. Teo and C. L. Kane,  Phys. Rev. B {\bf 89} , 085101 (2014).

\bibitem{yoreg} E. Sagi and Y. Oreg,  Phys. Rev. B {\bf 92}, 195137 (2015).

\bibitem{yan} Z. Yan, B. Li, X. Yang and S. Wan, Nat. Sci. Rep. {\bf 5}, 16197 (2015).

\bibitem{fulga} I. C. Fulga and M. Maksymenko, arxiv:1508.02726.

\bibitem{ezhao} E. Zhao, arxiv: 1603.08822.

\bibitem{diehl} S. Diehl {\it et.al}, Nat. Phys. {\bf 7}, 971-977 (2011).

\bibitem{goldman} N. Goldman and J. Dalibard, Phys. Rev. X, {\bf 4}, 031027 (2014).

\bibitem{Isingpaper}Manuscript under preparation.


\bibitem{SODWth} T. Das, Phys. Rev. Lett. {\bf 109}, 246406 (2012).

\bibitem{SODWexp} C. Brand {\it et. al}, Nat. Comm., {\bf 6}, 8118 (2015).

\bibitem{IsingSOC} J. M. Lu, O. Zeliuk, I. Leermakers, N. F. Q. Yuan, U. Zeitler, K. T. Law, J. T. Ye, Science {\bf 350}, 1353 (2015); X. Xi, Z. Wang, W. Zhao, J. Park, K. T. Law, H. Berger, L. Forro, J. Shan, K. F. Mak, Nat. Phys. 12, 139 (2016); B. T. Zhou, N. F.Q. Yuan, H. L. Jiang, K. T. Law, Phys. Rev. B {\bf 93}, 180501 (2016), G. Sharma, S. Tewari, arXiv:1603.08909.

\bibitem{Roth}  A. Roth, C. Brane, H. Buhmann, L. W. Molenkamp, J. Maciejko, X.L. Qi, and S.C. Zhang, Science {\bf 325}, 294 (2009).

\bibitem{gusev} G. M. Gusev,  E. B. Olshanetsky,  Z. D. Kvon,  A. D. Levin,  N. N. Mikhailov, and  S. A. Dvoretsky,   Phys. Rev. Lett {\bf 108}, 226804 (2012).


\end{thebibliography}
\end{document}